\documentclass[a4paper,10pt,twocolumn,english]{article}

\usepackage[english]{babel}
\usepackage[utf8]{inputenc}
\usepackage[T1]{fontenc}
\usepackage{booktabs} 
\usepackage[hidelinks]{hyperref}

\usepackage{graphicx}						
\usepackage{caption}			
\usepackage{subcaption}						
\usepackage{color}							

\usepackage{tikz}
\tikzset { domaine/.style 2 args={domain=#1:#2} }
\usepackage{pgfplots}
	\usetikzlibrary{shapes.misc}

\usepackage{amsmath, amssymb, amsfonts,mathtools}
\usepackage{amsthm}

\usepackage{flushend}

\title{\vspace{-.8cm}Blended smoothing splines on Riemannian manifolds
\vspace{-.2cm}}

\author{
	Pierre-Yves Gousenbourger$^\star$, Estelle Massart$^\star$ and P.-A. Absil$^\star$\\
	\footnotesize $^\star$ Universit\'e catholique de Louvain - ICTEAM Institute, 
	\footnotesize B-1348 Louvain-la-Neuve, Belgium \vspace{-.5cm} } 

\date{\empty}

	\usepackage[top=1.5cm, left=1.5cm, right=1.5cm, bottom=1.5cm]{geometry}

	\renewenvironment{abstract}{\bf\small {\em\ Abstract --}}{}

\definecolor{darkgreen}{rgb}{0,0.6,0.1}
\definecolor{darkblue}{rgb}{0,0,0.7}
\definecolor{mygreen}{rgb}{0.3,0.8,0}
\definecolor{myblue}{rgb}{0,0.5,1}
\definecolor{myred}{rgb}{1,0,0}
\definecolor{orange}{rgb}{1,0.7,0.2}
\definecolor{mypink}{rgb}{0.8,0.2,0.6}

\newcommand{\av}			{\mathrm{av}}

\newcommand{\R}				{\mathbb{R}}							
\newcommand{\M}				{\mathcal{M}}							
\newcommand{\C} 			{\mathcal{C}}							
\renewcommand{\d}     		{\mathrm{d}}							

\newcommand{\Exp}		[2]	{\mathrm{Exp}_{{#1}}\left({#2}\right)}	
\newcommand{\Log}		[2]	{\mathrm{Log}_{{#1}}\left({#2}\right)}	

\newcommand{\bspline}		{\mathbf{B}}							

\newcommand{\eg}{{e.g.}}

\newcommand{\etal}{{\emph{et al.}}}

\tikzset{
	cross/.style={
		cross out, 
		draw=black,
		minimum size=2*(#1-\pgflinewidth), 
		inner sep=0pt, 
		outer sep=0pt,
		line width=1.5pt,
		},
	cross/.default={4pt},
	bezier/.style={
		black, solid, line width=1pt
	},
	blending/.style={
		myblue, dash dot, line width=1pt
	},
	blOne/.style={
		orange, loosely dotted, line width=1pt
	},
	blTwo/.style={
		mygreen, densely dotted, line width=1pt
	},
	TS/.style={
		mypink, dashed, line width=1pt
	},
	Opt/.style={
		gray,
	},
}

\pgfdeclarelayer{bg}    
\pgfdeclarelayer{fg}    
\pgfsetlayers{bg,main,fg}  

\begin{document}
	\maketitle

	\begin{abstract} We present a method to compute a fitting curve
	$\bspline$ to a set of data points $d_0,\dots,d_m$ lying on a manifold 
	$\M$. That curve is obtained by blending together Euclidean
	B\'ezier curves obtained on different tangent spaces. The method
	guarantees several properties among which $\bspline$ is $\C^1$ and 
	is the natural cubic smoothing spline when $\M$ is the Euclidean space.
	We show examples on the sphere $S^2$ as a proof of concept.\footnote{
	This work was supported by 
	the Fonds de la Recherche Scientifique -- FNRS and the Fonds Wetenschappelijk Onderzoek -- Vlaanderen under 
	EOS Project no 30468160 
	and by ``Communauté française de Belgique - Actions de Recherche Concertées''.
	It also uses the toolbox Manopt~\cite{Boumal2014}.}
	\end{abstract}

	\section{Introduction}
	\label{sec:introduction}

	We address the problem of curve fitting on a Riemannian manifold $\M$.
	From a set of data points $d_0,\dots,d_m \in \M$ associated with times
	$t_0,\dots,t_m$ on a given time-interval $[0,n]$, we seek 
	a $\C^1$ curve $\bspline: [0,n] \to \M$ that is ``sufficiently straight'',
	while approximating ``sufficiently well'' the data points at the given 
	times.
	
	Curve fitting on manifold appears in several applications where denoising
	or resampling time-dependent data is required. For instance, in
	Arnould \etal~\cite{Arnould2015}, the evolution of an organ is
	observed by interpolating several contours of a tumoral tissue on a
	shape manifold. Regression is also of interest in problems where
	3D rigid rotations of objects are involved, as in motion planning
	of rigid bodies or in computer graphics~\cite{Park2010}. In that case,
	the manifold would be the special orthogonal group $\mathrm{SO}(3)$.
	
	A widely used strategy to address the fitting problem in general 
	is to encapsulate the fitting and straightness constraints in a
	single optimization problem
	\begin{equation}
		\label{eq:E}
		\min_{\gamma \in \Gamma} E_\lambda(\gamma) 
			\coloneqq \int_{t_0}^{t_m} \left\|\frac{\mathrm{D}^2 \gamma(t)}{\mathrm{d}t^2}
			\right\|_{\gamma(t)}^2 \mathrm{d}t +  \lambda \sum_{i=0}^m \d^2(\gamma(t_i),d_i),
	\end{equation}
	where $\Gamma$ is an admissible set of curves $\gamma$ on $\M$, 
		$\frac{\mathrm{D}^2}{\mathrm{d}t^2}$ is the (Levi-Civita) second covariant derivative, 
		$\|\cdot\|_{\gamma(t)}$ is the Riemannian metric at $\gamma(t)$, and
		$\d(\cdot,\cdot)$ is the Riemannian distance. 
	The parameter $\lambda$ permits to strike the balance between the regularizer 
	$\int_{t_0}^{t_m} \|\frac{\mathrm{D}^2 \gamma(t)}{\mathrm{d}t^2}\|_{\gamma(t)}^2 \mathrm{d}t$
	and the fitting term $\sum_{i=0}^m \d^2(\gamma(t_i),d_i)$.
	
	This problem has been tackled in different ways in the past few years.
	We cite for instance Samir \etal~\cite{Samir2012} that approached
	the solution of problem~\eqref{eq:E} with a manifold-valued 
	steepest-descent method	on an infinite dimensional Sobolev 
	space equipped with the Palais-metric. In Boumal \etal~\cite{Boumal2011},
	the search space is reduced to the product manifold $\M^M$, 
	as the curve $\bspline$	is discretized in $M$ points, 
	and the covariant derivative from~\eqref{eq:E} is approached with 
	finite differences on manifolds. A technique for regression based on
	unwrapping and unrolling has been recently proposed by Kim \etal~\cite{Kim2018}.
	Finally, we mention Lin \etal~\cite{Lin2017}, who proposed a
	polynomial regression technique based on projections on tangent spaces.
	
	The limit case when $\lambda \to \infty$ concerns interpolation. We
	cite here several works that solve this problem by means 
	of B\'ezier curves~\cite{Arnould2015,Absil2016}.
	In those works, the search space $\Gamma$ is reduced to composite
	cubic B\'ezier splines $\bspline$ and the optimality of~\eqref{eq:E} is guaranteed
	only when $\M = \R^r$. However, the main advantages of these methods
	are twofold: \emph{(i)} the search space is drastically reduced to the
	so-called \emph{control points} of $\bspline$ (see, \eg,~\cite{Farin2002}
	for an overview on B\'ezier curves); \emph{(ii)} they are
	very simple to implement on any Riemannian manifold, as only two
	objects are required: the Riemannian exponential and the Riemannian 
	logarithm, while most of the other techniques require a gradient or
	heavy computations of parallel transportation.
	
	Our method aims to extend these works to fitting, and is extensively
	described in~\cite{Gousenbourger2018} for the case where $m=n$. 
	We build several polynomial pieces by solving the 
	problem~\eqref{eq:E} on carefully chosen tangent spaces, and then 
	blend together these curves in such a way that $\bspline$ 
	\emph{(i)} is differentiable, \emph{(ii)} is the natural cubic smoothing
	spline when $\M$ is a Euclidean space, \emph{(iii)} interpolates
	the data points if $m = n$ when $\lambda \to \infty$.
	Furthermore, we assess that
	the method is easy-to-use, as \emph{(iv)} it only requires the knowledge 
	of the Riemannian exponential and the Riemannian logarithm on $\M$;
	\emph{(v)} the curve can be stored with only $\mathcal O(n)$ tangent vectors;
	and, finally, \emph{(vi)} given this representation, computing $\gamma(t)$ 
	at $t \in [0,n]$ only requires $\mathcal O(1)$ exponential and logarithm evaluations.
	
	We present here the above-mentioned method and give results for
	fitting on the sphere $\mathrm{S}^2$. We refer to~\cite{Gousenbourger2018}
	for more details and for the proof of the six properties.

	\begin{figure}[t!]
		\centering
		\begin{tikzpicture}[scale=.85]

\begin{pgfonlayer}{bg}
	\begin{scope}[shift={(0,0)}]
		\coordinate (Mbl) at (0,0);				
		\coordinate (Mtl) at (-1,2);
		\coordinate (Mtr) at (6,2);
		\coordinate (Mbr) at (7,-0.5); 
		
		\path (Mbr) to[bend left =20] node[pos=0.8](MD1){} (Mtl);	
		\path (Mbr) to[bend left =30] node[pos=0.4](MD2){} (Mtr);	
		
		\path (Mbl)+(0,0.5) to node[pos=0.15](MP1){} (Mtr);			
		\path (Mbr)+(0,0.5) to node[pos=0.15](MP2){} (Mtl);			
		\path (MP1) to +(1,1) node(MBm){};							
		\path (MP2) to +(-1.5,1) node(MBp){};						

		\draw[line width=1pt,gray] 
			(Mbl) .. controls +(-0.25,1.5) .. (Mtl) 
				  .. controls +(3,0.75) and +(-3,0.5) .. (Mtr)
				  .. controls +(1,-1.5) .. (Mbr)
				   .. controls +(-3,0.5) and +(2,0) .. cycle;

		\draw[gray] (Mbr) node[anchor=north]{$\M$};

		\path
			(MP1) .. controls (MBm) and (MBp) .. 
				node[pos=0.35](L){}
				node[pos=0.45](R){}
				(MP2);

		\draw(MD1) node[left]  {$d_i$}; 					 \fill[myred] (MD1) circle(.08);
		\draw(MD2) node[above right,fill=white] {$d_{i+1}$}; \fill[myred] (MD2) circle(.08);
		
		\fill[white] (MP1) circle(.08); \draw[mygreen,line width=1pt] (MP1) circle(.08); \draw(MP1) node[gray,below]  {$p_i$};
		\fill[white] (MP2) circle(.08); \draw[mygreen,line width=1pt] (MP2) circle(.08); \draw(MP2) node[gray,below]  {$p_{i+1}$};
		
		\fill[white] (MBm) circle(.08); \draw[mygreen,line width=1pt] (MBm) circle(.08); \draw(MBm)+(-0.1,0) node[gray,above]  {$  b_i^+$};
		\fill[white] (MBp) circle(.08); \draw[mygreen,line width=1pt] (MBp) circle(.08); \draw(MBp) node[gray,above] {$  b_{i+1}^-$};
		
		\fill[white](L) circle(.1); \fill (L) circle(.05); \draw (L) circle(.1); \draw (L) node[below left]{$L(t)$};
		\fill[white](R) circle(.1); \fill (R) circle(.05); \draw (R) circle(.1); \draw (R) node[below right]{$R(t)$};
		
		\path[draw,gray] (L) to node[pos=0.4](B){} (R);
		\fill[myblue] (B) circle(.08);
		
		\path[draw,<-,>=stealth,dotted,myblue] (B) to node[pos=1.2]{$\bspline(t)$} +(0,-1.6);
		
	\end{scope}
\end{pgfonlayer}

\begin{pgfonlayer}{fg}
	\begin{scope}[shift={(-1,3)},rotate=15]
		\coordinate (TblL) at (0,0);					
		\coordinate (TtlL) at (-1,2);
		\coordinate (TtrL) at (3,2);
		\coordinate (TbrL) at (4,0); 
		
		\path (TbrL) to[bend left =20] node[pos=0.8](TD1){} (TtlL);		
		\path (TblL)+(0,0.5) to node[pos=0.15](TP1L){} (TtrL);			
		\path (TP1L) to +(0.7,0.8) node(TBmL){};						
		\path (TP1L) to +(1.8,0.8) node(TBpL){};						
		\path (TP1L) to +(2.7,-0.4) node(TP2L){};						

		\draw[fill=white,opacity=.7]
			(TblL) -- (TtlL) -- (TtrL) -- (TbrL) -- cycle;
		\draw[line width=1pt,gray]
			(TblL) -- (TtlL) -- (TtrL) -- (TbrL) -- cycle;
			
		\draw[gray] (TblL) node[anchor=north east]{$T_{d_i}\M$};

		\path[draw,dashed] 
			(TP1L) .. controls (TBmL) and (TBpL) .. 
				node[pos=0.4](TL){}
				(TP2L);

		\draw (TD1) node[cross,red]{};
		
		\draw(TP1L)++(-0.1,0.2) node[fill=white,above]  	{$\tilde p_i$}; 		\draw[line width=1pt] (TP1L) node[cross,mygreen]{};
		\draw(TBmL)++(0,0.2) 	node[fill=white,above]  	{$\tilde b_i^+$}; 		\draw[line width=1pt] (TBmL) node[cross,mygreen]{}; 
		\draw(TBpL)++(0.4,-0.0) node[fill=white,above] 		{$\tilde b_{i+1}^-$}; 	\draw[line width=1pt] (TBpL) node[cross,mygreen]{}; 
		\draw(TP2L)++(0,0.2) 	node[fill=white,above right]{$\tilde p_{i+1}$}; 	\draw[line width=1pt] (TP2L) node[cross,mygreen]{}; 
		
		\fill[white](TL) circle(.1); \fill (TL) circle(.05); \draw (TL) circle(.1);
		
	\end{scope}
\end{pgfonlayer}

\begin{pgfonlayer}{fg}
	\begin{scope}[shift={(4,3)},rotate=-15]
		\coordinate (TblR) at (-0.7,0);					
		\coordinate (TtlR) at (-1.7,2);
		\coordinate (TtrR) at (2.5,2);
		\coordinate (TbrR) at (3.5,0);

		\path (TbrR) to[bend left=0] node[pos=0.15](TD2){} (TtlR);		
		\path (TD2)  to +(-0.3,0.1) node(TP2R){};						
		\path (TP2R) to +(-0.9,0.8) node(TBpR){};						
		\path (TP2R) to +(-2.3,1) node(TBmR){};							
		\path (TP2R) to +(-2.7,0.1) node(TP1R){};						

		\draw[fill=white,opacity=.7]
			(TblR) -- (TtlR) -- (TtrR) -- (TbrR) -- cycle;
		\draw[line width=1pt,gray]
			(TblR) -- (TtlR) -- (TtrR) -- (TbrR) -- cycle;
			
		\draw[gray] (TtlR) node[anchor=south west]{$T_{d_{i+1}}\M$};

		\path[draw,dashed] 
			(TP1R) .. controls (TBmR) and (TBpR) .. 
				node[pos=0.4](TR){}
				(TP2R);

		\draw (TD2) node[cross,red]{};
		
		\draw(TP1R)++(-0.1,0.2) node[fill=white,above]  	{$\hat p_i$}; 		\draw[line width=1pt] (TP1R) node[cross,mygreen]{};
		\draw(TBmR)++(0,0.2) 	node[fill=white,above]  	{$\hat b_i^+$}; 	\draw[line width=1pt] (TBmR) node[cross,mygreen]{};
		\draw(TBpR)++(0,0.0) 	node[fill=white,above right]{$\hat b_{i+1}^-$}; \draw[line width=1pt] (TBpR) node[cross,mygreen]{};
		\draw(TP2R)++(0,0.1) 	node[fill=white,above right]{$\hat p_{i+1}$}; 	\draw[line width=1pt] (TP2R) node[cross,mygreen]{};
		
		\fill[white](TR) circle(.1); \fill (TR) circle(.05); \draw (TR) circle(.1);
		
	\end{scope}
\end{pgfonlayer}

\path[->,>=stealth,draw,dotted,gray,line width=1pt] (TL) edge[bend left=20] (L);
\path[->,>=stealth,draw,dotted,gray,line width=1pt] (TR) edge[bend right=20] (R);

\path[draw,gray,line width=1pt] (MD1) edge[bend left=0] (TD1);
\path[draw,gray,line width=1pt] (MD2) edge[bend right=0] (TD2);

		\end{tikzpicture}
		\caption{The curve $\bspline(t)$ is made of natural cubic splines
			computed on different tangent spaces. The cubic splines can
			be obtained equivalently as B\'ezier curves, using a technique
			close to~\cite{Arnould2015}. They are then 
			blended together with carefully chosen weights.
			}
		\label{fig:blended}
	\end{figure}
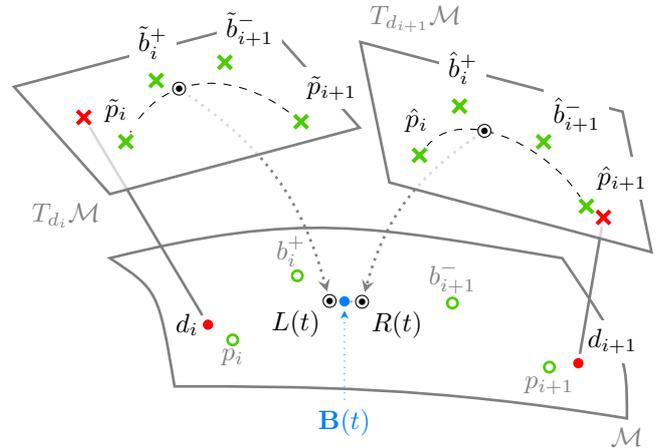
	
	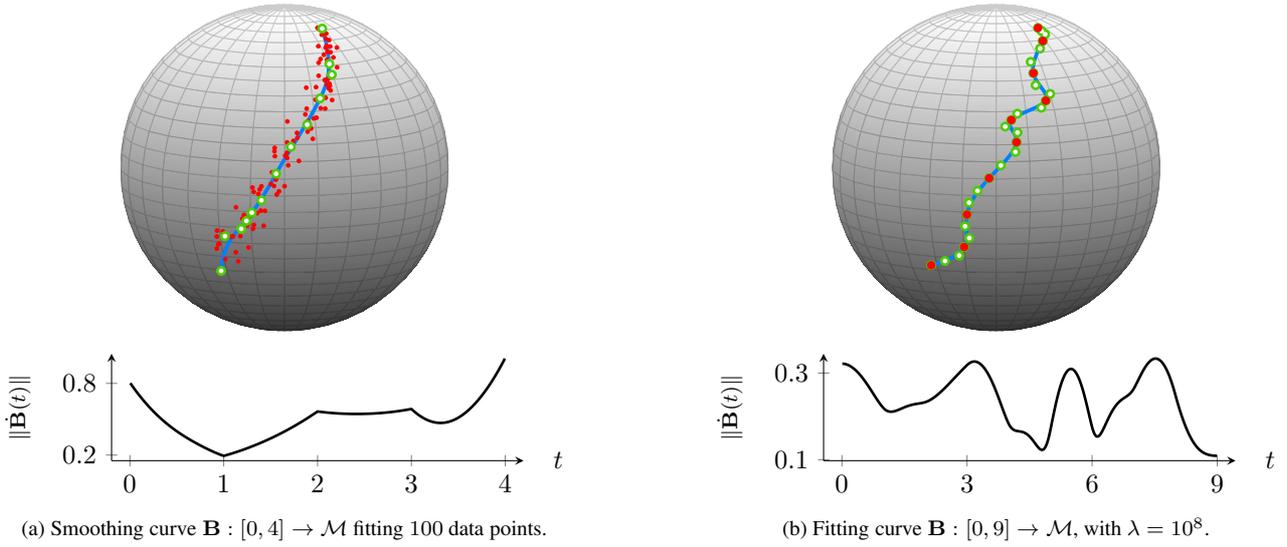
\begin{figure*}[t!]
		\centering
		\begin{subfigure}[t]{0.48\textwidth}
			\centering
			\begin{tikzpicture}[scale=.7]

\begin{axis}[%
	compat=1.6,					
	scale only axis,			
	unit vector ratio*=1 1 1,
	ticks = none,
	xmin=-1, xmax=1,			
	ymin=-1, ymax=1,			
	zmin=-.9, zmax=1,			
	view={-30}{10},				
	hide axis,
	]

	\addplot3 [surf, shader=faceted interp, z buffer=sort, colormap={bw}{gray(0cm)=(.2); gray(1cm)=(1)}, mesh/rows=31
	]
		table [x=x,y=y, z=z,col sep=comma]
		{pics/data-sphere.dat};

	\addplot3 [ myred,
				mark=*, 
				only marks,
				mark options={solid, scale=.5}]
		table [x=x,y=y, z=z]
		{pics/figure2/data-sphere_data.dat};

	\addplot3 [myblue,solid,line width=2pt]
		table [x=x,y=y, z=z]
		{pics/figure2/data-sphere_curve.dat};

	\addplot3 [ mygreen,
				mark=*, 
				only marks,
				mark options={solid, scale=1.2}]
		table [x=x,y=y, z=z]
		{pics/figure2/data-sphere_control.dat};
	\addplot3 [ white, 
				mark=*, 
				only marks,
				mark options={solid, scale=.5}]
		table [x=x,y=y, z=z]
		{pics/figure2/data-sphere_control.dat};
\end{axis}
			\end{tikzpicture}\\
			\begin{tikzpicture}
				\begin{axis}[
	axis x line = bottom,
	axis y line = left,
	enlarge x limits=0.05,
	enlarge y limits=0.05, 
	xlabel={$t$},
	ylabel={$\| \dot \bspline(t) \|$},
	x label style={font=\footnotesize},
	every axis x label/.style={
		at={(ticklabel* cs:1.05)},
		anchor=west,
	},
	y label style={font=\footnotesize},
	ytick={0.2,0.8},
	height=3cm,
	width=7cm
	]
	
	\addplot [bezier]
		table [x=x,y=y]
		{pics/figure2/data-sphere_speed.dat}
		;
		
\end{axis}
			\end{tikzpicture}
			\caption{Smoothing curve $\bspline:[0,4]\to\M$ fitting $100$ data points.}
			\label{fig:regression}
		\end{subfigure}
		\hfill
		\begin{subfigure}[t]{0.48\textwidth}
			\centering
			\begin{tikzpicture}[scale=.7]

\begin{axis}[%
	compat=1.6,					
	scale only axis,			
	unit vector ratio*=1 1 1,
	ticks = none,
	xmin=-1, xmax=1,			
	ymin=-1, ymax=1,			
	zmin=-.9, zmax=1,			
	view={-30}{10},				
	hide axis,
	]

	\addplot3 [surf, shader=faceted interp, z buffer=sort, colormap={bw}{gray(0cm)=(.2); gray(1cm)=(1)}, mesh/rows=31
	]
		table [x=x,y=y, z=z,col sep=comma]
		{pics/data-sphere.dat};

	\addplot3 [myblue,solid,line width=2pt]
		table [x=x,y=y, z=z]
		{pics/figure2/data-sphereFit_curve.dat};

	\addplot3 [ mygreen,
				mark=*, 
				only marks,
				mark options={solid, scale=1.2}]
		table [x=x,y=y, z=z]
		{pics/figure2/data-sphereFit_control.dat};
	\addplot3 [ white, 
				mark=*, 
				only marks,
				mark options={solid, scale=.5}]
		table [x=x,y=y, z=z]
		{pics/figure2/data-sphereFit_control.dat};
	
	\addplot3 [ myred,
				mark=*, 
				only marks,
				mark options={solid, scale=1}]
		table [x=x,y=y, z=z]
		{pics/figure2/data-sphereFit_data.dat};
\end{axis}
			\end{tikzpicture}\\
			\begin{tikzpicture}
				\begin{axis}[
	axis x line = bottom,
	axis y line = left,
	enlarge x limits=0.05,
	enlarge y limits=0.05, 
	xlabel={$t$},
	ylabel={$\| \dot \bspline(t) \|$},
	x label style={font=\footnotesize},
	every axis x label/.style={
		at={(ticklabel* cs:1.05)},
		anchor=west,
	},
	y label style={font=\footnotesize},
	xtick={0,3,6,9},
	ytick={0.1,0.3},
	height=3cm,
	width=7cm
	]
	
	\addplot [bezier]
		table [x=x,y=y]
		{pics/figure2/data-sphereFit_speed.dat}
		;

\end{axis}
			\end{tikzpicture}
			\caption{Fitting curve $\bspline:[0,9]\to \M$, with $\lambda = 10^8$.}
			\label{fig:fitting}
		\end{subfigure}
		\caption{The data points (red dots) are fitted by a $\C^1$ composite
			blended spline $\bspline(t)$ (blue). The blended spline is here
			represented as a B\'ezier curve conducted by its control points
			(green circles).
			}
		\label{fig:sphere}
	\end{figure*}
	
	\section{Method}
	\label{sec:method}
	\paragraph{Framework.} Consider a Riemannian manifold $\M$ and a set of $m+1$ data points 
	$d_0,\dots,d_m \in \M$ associated with parameters $t_0,\dots,t_m$ over
	an interval $[0,n]$.
	Our method relies on computations on tangent spaces. For this, we
	define the points $d(i)$, $i=0,\dots,n$, where $d(i) = d_{k_i}$ is the data point
	whose associated parameter $t_{k_i}$ is the closest to $t = i$. 
	We denote $T_{d(i)}\M$ its associated tangent space.
	Consider finally the search space $\Gamma$ from
	\eqref{eq:E} reduced to the space of $\C^1$ composite curves
	$$
		\bspline: [0,n] \to \M : f_i(t-i), \ i = \lfloor t \rfloor,
	$$
	where the functions $f_i: [i,i+1] \to \M$ are called 
	\emph{blended functions}. They are given by
	$$
		f_i(t-i) = \av[(L_i(t),R_i(t)),(1-w(t),w(t))],
	$$
	for $i = 0,\dots,n$ and where $\av[(x,y),(1-a,a)]$ is a Riemaniann weighted mean.
	The fitting technique we present here consists in computing the
	functions $L_i(t)$, $R_i(t)$ and choosing the weight function $w(t)$
	such that the six above-mentioned properties are met.

	\paragraph{Optimal curves.} The functions $L_i(t)$ and $R_i(t)$ are
	obtained as follows. We note $\tilde x = \Log{d(i)}{x}$ and
	$\hat x = \Log{d(i+1)}{x}$, the representation of the point $x\in\M$
	in the tangent spaces at $d(i)$ and $d(i+1)$ respectively. We define
	$L_i(t) = \Exp{d(i)}{\tilde \bspline(t)}$ and $R_i(t) = \Exp{d(i+1)}{\hat \bspline(t)}$,
	where $\tilde \bspline(t)$ is the natural cubic spline fitting the
	data points $\tilde d_0, \dots, \tilde d_m$ on $T_{d(i)}\M$, and
	accordingly for $\hat \bspline(t)$. Note that $\tilde \bspline(t)$
	(resp. $\hat \bspline(t)$) are therefore solutions of~\eqref{eq:E}
	on the corresponding tangent space.
	
	\paragraph{Riemannian averaging.} Finally, the choice of the weight
	function $w(t)$ is of high importance in order to meet the differentiability
	property. The weight function must thus be chosen such that
	$L_i(0) = f_i(0)$, $R_i(1) = f_i(1)$, $\dot L_i(0) = \dot f_i(0)$ and
	$\dot R_i(1) = \dot f_i(1)$. This is obtained for $w(1) = 1$, and
	$w(0) = w'(0) = w'(1) = 0$. Among all the possible weight functions,
	we choose $w(t) = 3 t^2 - 2 t^3$.
	
	The blending method is represented in Figure~\ref{fig:blended}.

	\section{Results}
	\label{sec:results}
	We show two examples on $\mathrm S^2$.
	Figure~\ref{fig:regression} presents a smoothing curve fitting
	$100$ noisy points at times $t_i \in [0,4]$ with $\lambda = 100$. Figure~\ref{fig:fitting}
	shows the fitting curve obtained for $10$ data points at times
	$t_i = i$, $i=0,\dots,9$, for $\lambda = 10^8$. We observe in both cases
	that the curve is $\C^1$ (property \emph{(ii)}) and that the data
	points are interpolated (property \emph{(iii)}) when $\lambda \to \infty$.
	Property \emph{(i)} is obtained by construction. Properties \emph{(iv-vi)}
	are shown and proved in~\cite{Gousenbourger2018}. Additionnal examples
	on the special orthogonal group $\textrm{SO}(3)$ or on the manifold of
	positive semidefinite matrices of size $p$ and rank $q$, $\mathcal S_+(p,q)$,
	are also provided in~\cite{Gousenbourger2018}.

	\end{document}